\documentclass[]{aa}
\usepackage{txfonts}
\usepackage{subfig}
\usepackage{upgreek}

\usepackage{graphicx}
\usepackage{multirow}
\usepackage{amsmath,amssymb,amsfonts}
\usepackage{mathrsfs}
\usepackage[title]{appendix}
\usepackage{xcolor}
\usepackage{textcomp}
\usepackage{manyfoot}
\usepackage{booktabs}
\usepackage{algorithm}
\usepackage{algorithmicx}
\usepackage{algpseudocode}
\usepackage{listings}
\usepackage{booktabs}
\usepackage[flushleft]{threeparttable}
\usepackage{threeparttablex}
\usepackage{tabularray}

\usepackage[colorlinks=true]{hyperref}

\hypersetup{
     colorlinks   = true,
     citecolor    = blue
}

\usepackage{orcidlink}

\begin{document} 

\renewcommand{\arraystretch}{1.5}

    \title{Decoding the jet of BL Lacertae using relativistic magneto-hydrodynamics}
    \author{
G.~F. Paraschos\inst{1,2,3} \orcidlink{0000-0001-6757-3098},
J.~A. Kramer\inst{4,3} \orcidlink{0009-0003-3011-0454},
I. Liodakis\inst{5,6,3}\orcidlink{0000-0001-9200-4006},
S.~G. Jorstad\inst{7,8}\orcidlink{0000-0001-9522-5453},
A.~P. Marscher\inst{7}\orcidlink{0000-0001-7396-3332},
I. Myserlis\inst{9}\orcidlink{0000-0003-3025-9497},
I. Agudo\inst{10}\orcidlink{0000-0002-3777-6182},
N.~R. MacDonald\inst{11,3}\orcidlink{0000-0002-6684-8691}
}
    \authorrunning{G.~F. Paraschos et al.}
    \institute{
$^{1}$Finnish Centre for Astronomy with ESO, University of Turku, 20014 Turku, Finland\\
$^{2}$Aalto University Metsähovi Radio Observatory, Metsähovintie 114, FI-02540 Kylmälä, Finland\\
$^{}$\ \email{gfpara@utu.fi}\\
$^{3}$Max-Planck-Institut f\"{u}r Radioastronomie, Auf dem H\"{u}gel 69, D-53121 Bonn, Germany\\
$^{4}$Anton Pannekoek Institute for Astronomy, University of Amsterdam, Science Park 904, 1098 XH, Amsterdam, The Netherlands\\
$^{5}$Institute of Astrophysics, Foundation for Research and Technology - Hellas, Voutes, 7110, Heraklion, Greece\\
$^{6}$Institute of Astrobiology, University of Crete, Heraklion, GR-70013, Greece\\
$^{7}$Institute for Astrophysical Research, Boston University, 725 Commonwealth Avenue, Boston, MA 02215, USA\\
$^{8}$Saint Petersburg State University, 7/9 Universitetskaya nab., St. Petersburg, 199034 Russia\\
$^{9}$Institut de Radioastronomie Millim\'{e}trique, Avenida Divina Pastora, 7, Local 20, E–18012 Granada, Spain\\
$^{10}$Instituto de Astrof\'{i}sica de Andaluc\'{i}a, IAA-CSIC, Glorieta de la Astronom\'{i}a s/n, E-18008 Granada, Spain\\
$^{11}$Department of Physics and Astronomy, University of Mississippi, University, Mississippi 38677, USA\\
}

   \date{Received -; accepted -}

\abstract{
Blazars are a highly variable subclass of active galactic nuclei, whose relativistic jet is pointed towards our line of sight at a small angle.
Their variability is often characterised by multi-band flares.
BL Lacertae (BL Lac), the namesake of a blazar subclass, recently exhibited the highest recorded linearly polarised optical flare.
We investigate the origin of this flare via very-long-baseline interferometry observations.
Our analysis shows that the sweeping, helical motion of the BL Lac jet, which is known to exhibit kink-like instabilities, can explain the observed flux density spike and polarisation angle rotation, as also confirmed by our state-of-the-art relativistic magneto-hydrodynamic simulations.
As a by-product of these simulations, we find that baryon loading of the jet is required to optimally replicate the observed jet morphology.
}


   \maketitle

\section{Introduction}\label{sec1}

Blazars are often characterised by extreme variability across the electromagnetic spectrum and by highly polarised emission due to accelerated particles in the magnetic field of the jet \citep{Blandford19, Hovatta19b}.
The exact origins of their emission processes are still an open question \citep[e.g.][]{Boula22}.
Recently, BL Lacertae, the archetypal blazar of the BL Lac sub-class, exhibited the highest linear polarisation peak ever recorded in optical wavelengths since the first observations of blazars in linearly polarised light ($m$) in the 1970s \citep{Agudo25}. 
The measured value of $m_\textrm{opt} = 47.6\%$ is the highest optical polarisation degree ever observed in a relativistic jet from a SMBH, given the fact that the maximum, theoretically possible fractional linear polarisation ($m=P/I$, with $P$ and $I$ denoting the linearly polarised flux density and total intensity, respectively) is of the order of $m_\textrm{max} = 70\%$ \citep{Rybicki79}. 
However, this is only true for a single perfectly ordered magnetic field direction. 
Incoherent averaging of multiple magnetic field orientations leads to a reduction of the maximum theoretical polarisation degree by a factor of $1/\sqrt{N}$, where $N$ is the number of different magnetic field direction in the line of sight \cite[e.g.,][]{Marscher14,Peirson2018}. 
The observed optical polarisation degree suggests $N=2$, which is consistent with a perfectly ordered magnetic field in the emission region, likely a jet component,  in combination with the extended  pc-scale jet emission.

In this work, we utilise the full power of very-long-baseline interferometry (VLBI) to resolve the compact emission region of BL Lac to scrutinise its structure and test scenarios of linear polarisation variability connected to this recent outburst.
If interpreted correctly, this variability offers a number of clues into the underlying blazar physics, by acting as a proxy of measurables, such as the relativistic Doppler boosting, brightness temperature, and magnetic field \citep[see e.g.][]{Liodakis21}.
Indeed, magnetic fields are of paramount importance when studying the origin of blazars, since it is well established by now that they are responsible for launching jets \citep{Blandford77, Blandford82, Tchekhovskoy11, Tchekhovskoy15}.
A direct probe of the magnetic field orientation is polarisation; VLBI is a particularly well-suited technique to determine the polarisation signal.
VLBI enables us to detect parsec-scale structural changes of the blazar in the ultimate vicinity of the central supermassive black hole \citep[SMBH;][]{Jorstad05, Paraschos22, Debbrecht26}, and even around its event horizon \citep[e.g.][]{EHT21a, EHT24a, EHT24b}, which can then be modelled \citep[e.g.][]{GomezMiller26}.
The nature of these structural changes is a well-researched topic.
Geometrical effects, such as bends in a twisted, rotating jet, amplified by Doppler boosting, have been proposed as a possible mechanism causing variability \citep{Raiteri21, Raiteri24}.
In this framework, a bright jet feature propagates downstream and interacts with stationary shocks \citep[e.g.][]{Marscher08, Marscher10}.
Another possibility is the propagation of magneto-hydrodynamic instabilities downstream of the jet \citep{Raiteri17a}; magnetic field turbulence can also explain the observed phenomenology (\citealt{Raiteri17b}; see also \citealt{Raiteri25} for a comprehensive review).

Observationally, such structural changes are often manifested in the form of new jet component ejections.
These components might be associated with extreme values of total intensity and polarisation, called flares.
Such flares are usually detectable across the electromagnetic spectrum, from $\gamma$-rays and optical wavelengths down to radio-waves.
Due to opacity effects, the higher frequency radiation reaches us first, whereas the lower frequency radio waves (used to create the VLBI images of the blazars) are trailing.
In such a sequence of images, if the apparent changes to the jet structure become clear, the emergence of new jet features can then be tied to flares \citep{Liodakis20, Paraschos23, Paraschos25a}.

\begin{figure*}[!ht] 
	   \includegraphics[width=0.978\textwidth]{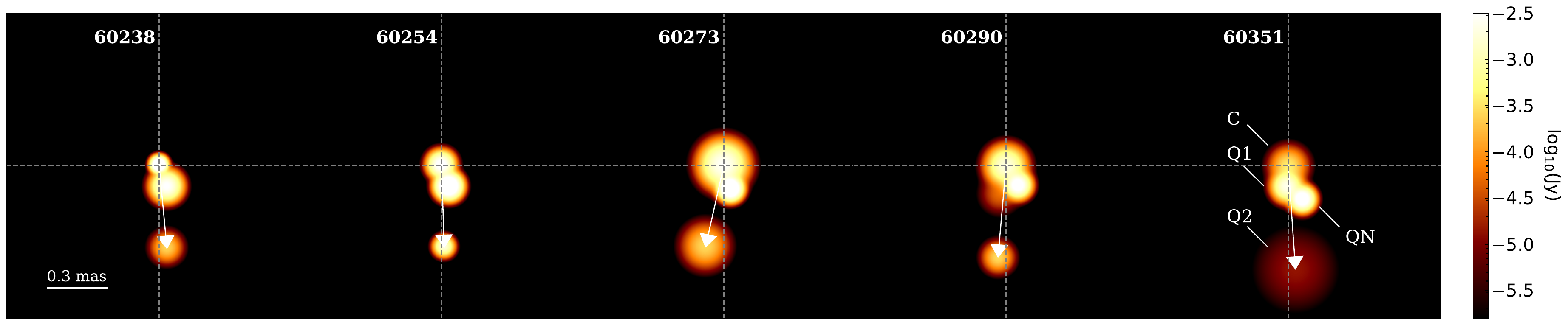}
        \includegraphics[trim={0cm 16cm 2.5cm 8cm}, clip, width=0.99\textwidth]{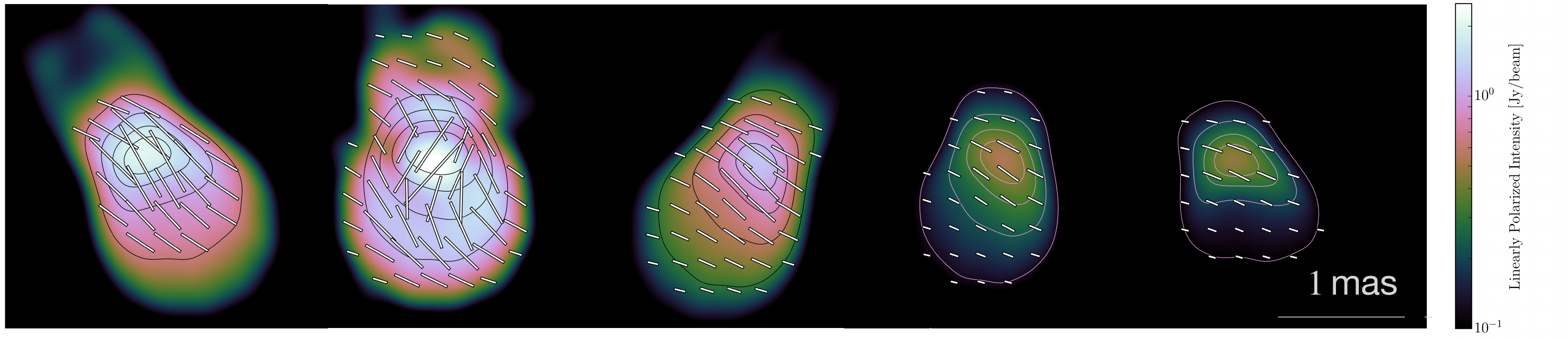}     
	\caption{
        Total intensity geometrical model fitting and linear polarisation RMHD simulations of the structure of BL Lac.
		\emph{Top panel}: From left to right, the panels correspond to MJD 60238, 60254, 60273, 60290, and 60351.
        Component Q2, denoted with the white arrow (see also Table~\ref{table:Params}), exhibits a swinging motion, between position angles 175$^\circ$ and $185^\circ$.
        At the time of the intensity peak in the optical flux, Q2 was aligned to our line of sight suggesting a connection to Doppler boosting.
        \emph{Bottom panel}: The logarithmically scaled, linearly polarised flux density is shown with the colour bar, while Stokes I is shown with the black contours (at 20\%, 40\%, 60\%, and 80\% of the individual epoch's linear intensity peak).
        The white sticks correspond to $\psi$.
        The simulations clearly display a similar behaviour to the VLBI observations.
        The rotation of the simulated jet mirrors both the initial Stokes I increase (when Q2 is pointing towards us), as well as the subsequent decrease (when the jet is pointing away from us).
        Similarly, $\psi$ exhibit a comparably smooth rotation to the observations, confirming our sweeping motion assumption.
            }
	\label{fig:MultiT} 
\end{figure*}

\begin{figure}[!ht]
	\centering
	   \includegraphics[width=0.5\textwidth]{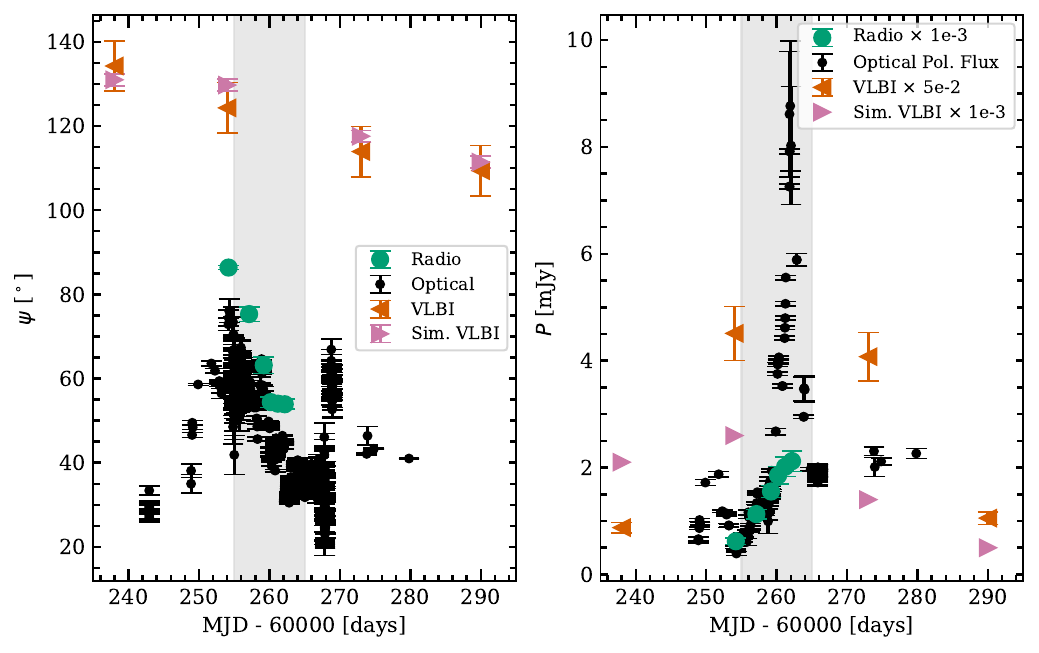}
	\caption{
        Optical, radio, and VLBI lightcurves of BL Lac.
        Left: Presented here is $\psi$ of BL Lac in optical (black) and radio (green) points during the IXPE observing window of MJD 60235 to 60290.
        The orange (purple) markers denote the VLBI measurements (simulations).
        Right: Displayed here is $P$ versus time.
        The noted scaling factors in the label are implemented to assist the reader with comparing the shape of the measurables with each other.
        The markers and colours are similar to the left panel.
            }
	\label{fig:LCs} 
\end{figure}

\begin{figure}[!ht]
	\centering
	   \includegraphics[width=0.5\textwidth]{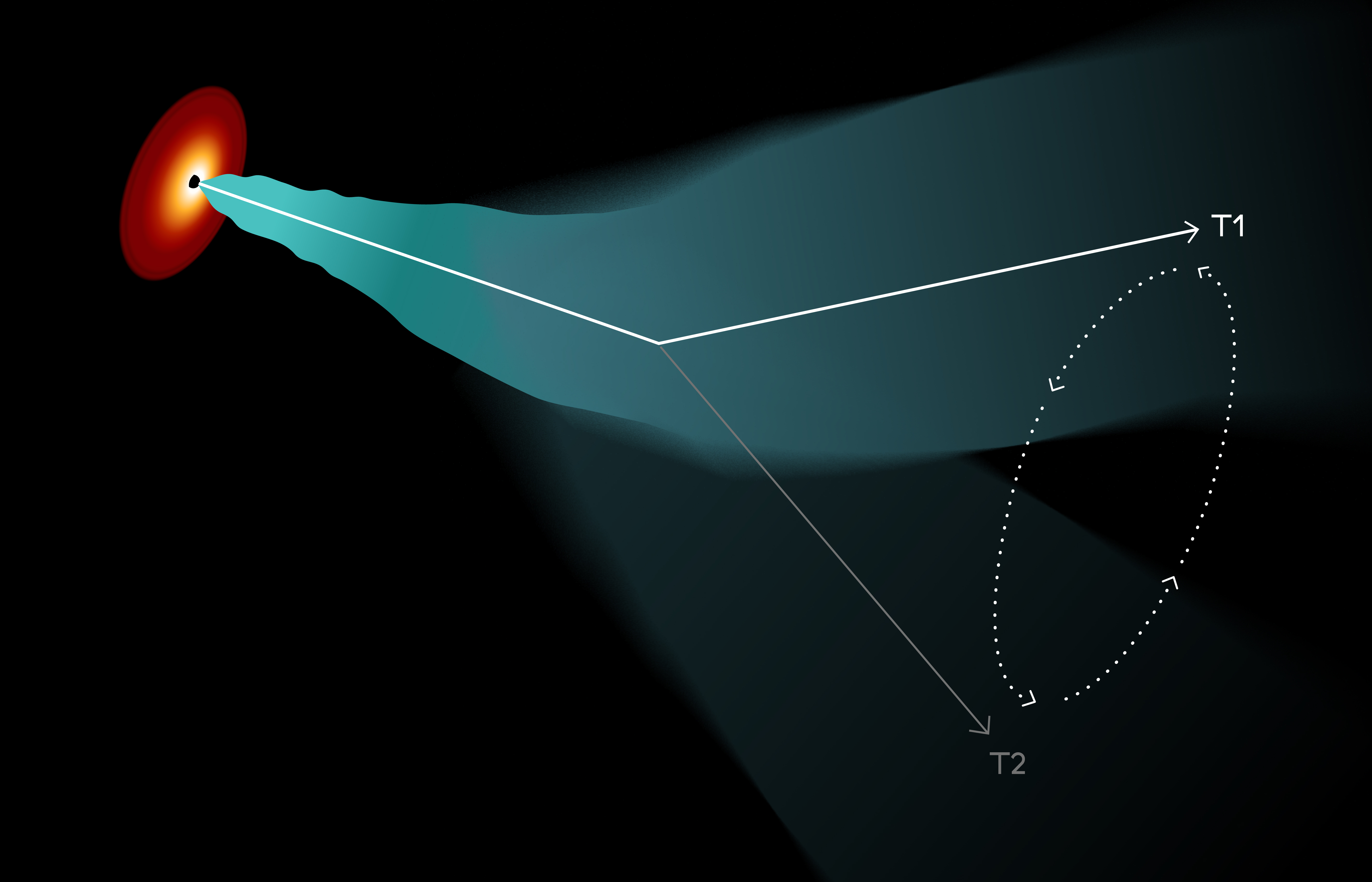}
	\caption{
        Cartoon of the BL Lac jet.
        The sweeping motion of the jet can explain the observed $\psi$ rotation and $P$ enhancement, as shown by our simulations and analytic calculations.
        The T1 and T2 markers indicate the times when the jet is at its westernmost and easternmost position of its trajectory with regard to the distant observer, respectively.
            }
	\label{fig:cartoon} 
\end{figure}

\section{Observations and evolution of BL Lac} \label{sec:obs}

BL Lac was monitored as part of the Imaging X-ray Polarimetry Explorer (IXPE) campaign \citep{Weisskopf22} -- interestingly no X-ray polarisation detection was reported, which, along with the low millimetre polarisation detection ($\sim10\%$), is suggestive of synchrotron self-Compton processes governing the source, as shown in \cite{Liodakis25}.
Furthermore, it is also part of the \texttt{BEAM-ME} sample of blazars\footnote{\url{https://www.bu.edu/blazars/BEAM-ME.html}}, which is regularly observed by the Very Long Baseline Array (VLBA) at 43\,GHz.
In the time frame of MJD 60235 to 60290, BL Lac was observed by the aforementioned radio and X-ray programmes, exhibiting the record linear polarisation flare at optical wavelengths on MJD 60261-60262.75.
During the time frame of interest, the source was also observed five times by the VLBA, as shown in the top panel of Fig.~\ref{fig:MultiT}; the source's structural evolution in its historical context is described below.

BL Lac has been known to exhibit rapid variability and structural changes for decades \citep{Cohen14, Jorstad17}.
In the top panel of Fig.~\ref{fig:MultiT}, we showcase the temporal progression of the structure of the source in the time frame of interest.
Following \cite{Casadio21}, we assumed that the core of the jet is stationary and identified it with the northernmost component in each epoch.
In Table~\ref{table:Params}, each core component is labelled with the letter `C'.
Two more components are identified in the compact region (defined as the inner $0.5\,\textrm{mas}$ area) of BL Lac, labelled as `Q1' and `Q2'.
A third component (`Q3') was used to model the extended, downstream emission of the jet (beyond the limits of the top panel of Fig.~\ref{fig:MultiT}).
In the last two epochs, a new component labelled as `QN' is ejected from the core region, perhaps associated with the detected flare.
Given its limited movement, Q1 might be the quasi-stationary component that has been identified in BL Lac in the past \citep{Cohen14, Jorstad22}.
On the other hand, Q2 and QN seem more likely to be associated with the flare, as discussed below.

Two major morphological characteristics are identified during the flaring state.
Firstly, the total flux of the components comprising the jet structure increased during the flare peak.
This flux density boost might be connected to Q2 being aligned to our line of sight, as it was undergoing a sweeping motion, while also moving away from the core region.
Furthermore, the flare culminated with the ejection of the new component QN, and the connection between flux density enhancement and new jet component ejection is well established in general \cite[e.g.][]{Savolainen02, Paraschos25b}, as well as for BL Lac specifically \cite[e.g.][]{Mutel90}.
Interestingly, hints of this helical motion are also present when examining the position angles of QN in the latter two epochs.
This is in line with past studies of BL Lac, which have revealed a connection between component ejections and flares \citep{Marscher08}, as well as the presence of such helical structures within the jet \citep{Denn00, Casadio21, Kim23}. 
An additional approach to gain further insights into the cause of the flare is by studying $P$ and the associated magnetic field of the component of interest component Q2.
As shown in Fig.~\ref{fig:LCs}, the VLBI polarisation angle ($\psi$) of that component rotates smoothly with time, similar to the radio and optical measurements. 
Furthermore, the VLBI $P$ increases between the first and second VLBI epoch and decreases afterwards, mirroring the optical and radio measurements.

Assuming that the flare in $P$ was caused by the sweeping motion of the Q2 component, we used the analytical approach by \citep{Hagen-Thorn2008} (see Eq. 1) to estimate the relation between the flux density and the viewing angle, expressed in terms of the Doppler factor $\delta$, which in turn is a function of the bulk Lorentz factor $\Gamma$.
This assumption is well motivated; if the jet consists of an unpolarised and a polarised part (Q2 being the latter), their superposition exactly reproduces the measured $m$ as explained in the introduction.
We find that for a $\Gamma=4.5$ \citep{Cohen15}, a viewing angle change of $\Delta\theta\approx15^\circ$ is sufficient to reproduce the observed $P$ changes.
While the large-scale viewing angle to the line of sight of BL Lac is $\theta<10^\circ$ \citep[e.g.][]{Casadio21}, localised, twisted patterns of the jet might exhibit larger viewing angles such as that one. 
This also fits our observation: Q2 moves a projected $\Delta\theta\leq10.7^\circ$ in the sky (see also Table~\ref{tab:angle_data}).
Our interpretation is also in line with the findings by \cite{Casadio21}; in Fig. 9 of their work, the authors show that at the position of Q2 the jet collimation decreases rapidly, becoming significantly wider.
Therefore, a scenario in which the alignment of the emerging component QN with the sweeping Q2 further downstream can explain the record-setting linear polarisation flare in BL Lac.

\section{Simulations of the BL Lac jet structure}

To further explore the sweeping motion scenario, we used state-of-the-art 3D RMHD simulation of a cylindrical non-axisymmetric jet to explore the necessary physical conditions that result in the observed $P$ and $\psi$ morphology \citep{Mignone07}, using the aforementioned, analytically calculated boundary conditions. 
Within this RMHD framework, a hybrid fluid-particle approach can be used: non-thermal electron populations are represented via Lagrangian particles integrated into the fluid simulation \citep{Vaidya18}. 
However, the final emission maps here are based on fluid only simulations, as~\cite{Kramer24} find them to result in the same synthetic synchrotron structure when leaving out particle acceleration terms.
The final simulation stage maps are shown in the left panel of Fig.~2 of ~\cite{Kramer21} and the pages following; we base our results in this paper on ray-traced linearly polarised emission maps of non-thermal particle attributes using the PLUTO code, our own numerical scheme for `electron painting' and the radiative transfer code RADMC-3D~\citep{Kramer21, Paraschos24b, Kramer24}. 
The simulations were performed in a Cartesian computational domain spanning $[-8,8]\times[0,40]\times[-8,8]$ in $(x,y,z)$ on a grid resolution of 320$\times$400$\times$320. 
Here we note that we do not achieve sufficient resolution to study turbulences. 
For a more complete view on quasi-periodic signals due to kink modes, see, for instance, ~\citet{Dong20, Bodo21, Acharya21, Hu25}. 
Furthermore, the synchrotron self-Compton (SSC) cooling was not included in the present synchrotron-only radiative-transfer calculations. 
However, because our simulations focus on the parsec-scale jet, where synchrotron emission is expected to dominate, we do not expect neglecting SSC to significantly affect the inferred $\psi$ morphology.
More information on our simulation can be found in Appendix~\ref{app:num}.

The resulting $P$ (colour) and total intensity (contours) images of a rotating blazar, representative of the swinging Q2 component, are shown in the bottom panel of Fig.~\ref{fig:MultiT}.
We find that, similar to the observations, the Stokes I flux density increases while the jet is pointing towards our line of sight and then decreases while the jet is pointing away. 
The simulations reproduce the observations most optimally for a jet sweep of $\Delta\theta\sim7^\circ$, adhering to the upper limit imposed by the observations. 
While the observations are variable in time, we chose our simulations to be variable in inclination.
In addition, smooth $\psi$ rotations can be caused by shocks compressing the field and geometric effects like jet bending or precession. 
We note that our simulations are consistent and steady state throughout the simulation, with a focus on the jet spine.
Figure~\ref{fig:LCs} demonstrates the excellent agreement between the observed and simulated values for both $\psi$ and $P$.
In our case, we observe the recollimation shock associated with the radio core while rotating the jet to some extent. 
Our simulations assume that $\psi$ are perpendicular to the magnetic field orientation in the optically thin regime. 

Finally, we explored the parameter space of the ratio between the hadrons and leptons in the jet when looking at the synthetic synchrotron emission in the observer frame.
We defined the plasma parameter $n$ as the ratio between protons and positrons in the jet.
The different realisations are shown in Fig.~\ref{fig:SimP} and the resulting $\psi$ orientations in Fig.~\ref{fig:SimE}.
Our investigation shows that the optimal match between observations and simulations is achieved when protons outnumber positrons by a factor of $n\geq100$.
Conversely, higher positron numbers fail to induce the observed $\psi$ orientation; the simulated $\psi$ remain constant regardless of the orientation of the jet, consistent with the dependence of the Faraday rotation coefficient on the plasma composition~\citep[Eq.~C.2;][]{MacDonald21}.
We note with interest that decreasing the number of proton around the time of the flare would increase the simulated $P$, matching the observations even more precise.
The accelerated protons would then start escaping the emission region (the size of $\Theta_\textrm{FWHM}$; $c$ is the speed of light), characterised by a light crossing time of $t_\textrm{cr} = \Theta_\textrm{FWHM}/c$, which yields $t_\textrm{cr}\sim$ few weeks.
This lower limit time frame matches the cadence of our VLBI observations, naturally explaining the quick return to quiescence after the flare peak.

Overall, our work paints a consistent picture: through simulations and analytical calculations, we have shown that the most prominent $P$ optical flare ever observed in a blazar can be effectively described by the sweeping motion of its jet, which is composed of a significant hadronic component.
A cartoon of this phenomenology is shown in Fig.~\ref{fig:cartoon}.

\section{Discussion and conclusions}

Interestingly, our results position BL Lac as a potential candidate for astrophysical neutrino emission.
An increasing number of recent studies have shown that blazars are potential neutrino producers \citep[e.g.][]{Plavin21, Hovatta21, Kouch24, Kouch25b, Paraschos25c}.
A prerequisite condition for neutrino creation is the existence of protons within the astrophysical jet, as shown in works by \cite{Petropoulou20}, \cite{Mastichiadis21}, \cite{Liodakis22a}, \cite{Stathopoulos26}, \cite{Traianou26}, among others, because this hadronic component is the catalyst for proton-photon or photomesonic interactions, which create neutrinos in the astrophysical context.
The expected neutrino output from the parsec-scale region studied here is discussed in Appendix~\ref{app:neutrino}: while the inferred proton kinetic luminosity is comfortably sub-Eddington ($L_\mathrm{p} \approx 5\times10^{-3}\,L_\mathrm{Edd}$), the photomeson efficiency of this region is low, implying that detectable neutrino emission would require the hadronic component to persist in more compact regions closer to the central engine.
Should such conditions be realised, then, with the commissioning of new neutrino observatories \citep{KM3Net25} and the expanding capabilities of existing ones \citep{Aartsen21}, a spatio-temporal association of an astrophysical neutrino with BL Lac may become possible in the future. 
Such an association would then probe these inner, more compact regions of the jet, and provide theoretical support for a prominent hadronic component within neutrino-coincident blazars.

In conclusion, in this work, we have investigated the mechanism that caused the most prominent $P$ optical flare ever in a blazar.
We found that the helical swing of the approaching jet reproduces the observed increase in Stokes I flux density and $P$, as well as the smooth rotation of $\psi$, by comparing our VLBI observations to RMHD simulations as a first order approximation.
Our results are further supported by analytical calculations, which indicate that a mild viewing angle change can reproduce the observed phenomenology.
Ultimately, our work underscores the major impact of relativistic jet geometry on multi-wavelength variability, suggesting that many extreme blazar phenomena may be driven as much by geometric perspective as by intrinsic energetic changes.

\begin{acknowledgements}
GFP and JAK contributed equally to this work.
We thank the anonymous reviewer for the constructive comments; we also thank S.~I. Stathopoulos and L.~C. Debbrecht for fruitful discussions.
This research is supported by the European Research Council advanced grant “M2FINDERS - Mapping Magnetic Fields with INterferometry Down to Event hoRizon Scales” (Grant No. 101018682). 
JAK is supported by a European Research Council Synergy Grant `BlackHolistic' grant No. 1010716.
IL was funded by the European Union ERC-2022-STG - BOOTES - 101076343. Views and opinions expressed are however those of the author(s) only and do not necessarily reflect those of the European Union or the European Research Council Executive Agency. Neither the European Union nor the granting authority can be held responsible for them. 
This research was partially funded by the Deutsche Forschungsgemeinschaft (DFG, German Research Foundation) as part of the DFG Research Unit FOR5195 – project number 443220636.
Some of the data are based on observations collected at the Observatorio de Sierra Nevada; which is owned and operated by the Instituto de Astrof\'isica de Andaluc\'ia (IAA-CSIC); and at the Centro Astron\'{o}mico Hispano en Andalucía (CAHA); which is operated jointly by Junta de Andaluc\'{i}a and Consejo Superior de Investigaciones Cient\'{i}ficas (IAA-CSIC). The Perkins Telescope Observatory, located in Flagstaff, AZ, USA, is owned and operated by Boston University. 
The Liverpool Telescope is operated on the island of La Palma by Liverpool John Moores University in the Spanish Observatorio del Roque de los Muchachos of the Instituto de Astrofisica de Canarias with financial support from the UKRI Science and Technology Facilities Council (STFC) (ST/T00147X/1). 
This research has made use of data from the RoboPol program, a collaboration between Caltech, the University of Crete, IA-FORTH, IUCAA, the MPIfR, and the Nicolaus Copernicus University, which was conducted at Skinakas Observatory in Crete, Greece. The data in this study include observations made with the Nordic Optical Telescope, owned in collaboration by the University of Turku and Aarhus University, and operated jointly by Aarhus University, the University of Turku, and the University of Oslo, representing Denmark, Finland, and Norway, the University of Iceland and Stockholm University at the Observatorio del Roque de los Muchachos, La Palma, Spain, of the Instituto de Astrofisica de Canarias. The data presented here were obtained in part with ALFOSC, which is provided by the Instituto de Astrof\'{\i}sica de Andaluc\'{\i}a (IAA) under a joint agreement with the University of Copenhagen and NOT. The Submillimetre Array (SMA) is a joint project between the Smithsonian Astrophysical Observatory and the Academia Sinica Institute of Astronomy and Astrophysics and is funded by the Smithsonian Institution and the Academia Sinica. Mauna Kea, the location of the SMA, is a culturally important site for the indigenous Hawaiian people; we are privileged to study the cosmos from its summit. The research at Boston University was supported in part by National Science Foundation grant AST-2108622, NASA Fermi Guest Investigator grant 80NSSC23K1507, NASA NuSTAR Guest Investigator grant 80NSSC24K0547, and NASA Swift Guest Investigator grant 80NSSC23K1145. This work was supported by NSF grant AST-2109127. We acknowledge funding to support our NOT observations from the Finnish Centre for Astronomy with ESO (FINCA), University of Turku, Finland (Academy of Finland grant nr 306531).
\end{acknowledgements}

\bibliographystyle{aa} 
\bibliography{aanda}

\begin{appendix}

\section{Multi-wavelength data}

The radio and optical data are part of the follow up campaign for the IXPE observation. 
More details can be found in \cite{Agudo25}. 
Here we provide a short description: The mm-radio observations at 225.5\,GHz were taken as part of the SMA Monitoring of AGNs with Polarisation (SMAPOL) program \cite{Myserlis25}. 
The observations were taken from 06-11-2023 (MJD~60254.19902) until 14-11-2023 (MJD~60262.17636) showing a median of $P\sim1\,$Jy, while $\psi$ drifted from 86$^\circ$ to 53$^\circ$. 
During the IXPE observation, BL Lac had the brightest mm-radio flare in the past 40 years \cite{Agudo25}. 

The optical observations use in this paper were taken in the R band from  eight different observatories, namely, the Belogradchik Observatory, Calar Alto Observatory, Nordic Optical Telescope, Liverpool telescope, LX-200, Perkins Telescope, Sierra Nevada Observatory, and the Skinakas observatory. 
$P$ climbed from $\sim1\,$mJy to $\sim10\,$mJy with a median of $\sim6\,$mJy; $\psi$ followed a similar pattern to the mm-radio going from about 70$^\circ$ to 30$^\circ$ during the IXPE observation. 
At the same time, there was a $\sim1$ magnitude flare that coincided with mm-radio flare and the sharp rise in $P$.

Finally, the 43\,GHz, publicly available VLBI data used here were calibrated and imaged as described in \cite{Jorstad17} and \cite{Weaver22}.
Our image reconstruction is described below.

\section{Geometrical model fitting}

\begin{table*}[ht] 
\centering
\begin{talltblr}[
  caption = {Summary of the component parameters},
  label = {table:Params},
  note{a} = {Flux density; $\Delta F0\sim\pm$10\%},
  note{b} = {Full Width at Half Maximum based on Gaussian fit; $\Delta\textrm{FWHM}\sim\pm$20\%},
  note{c} = {Cartesian position; $\Delta\rho\sim\pm0.04$\,mas},
  note{d} = {Linear polarisation flux density; $\Delta P\sim\pm$15\%},
  note{e} = {Polarisation angle; $\Delta\psi\sim\pm7^\circ$},
]{
  colspec = {ccccccc},
  hline{1, Z} = {0.08em}, 
  hline{2} = {0.05em},
}
ID & Obs. [MJD - 60000] & F0$^a$ [Jy] & FWHM$^b$ [mas] & Position$^c$ [mas, mas] & $P$$^d$ [Jy] & $\psi$$^e$ [$^\circ$]\\
\hline\hline
 C  &  238 &  6.362  &  0.04 &  [0.00, 0.00]   &  0.42 &  15 \\
 Q1 &  238 &  5.570  &  0.08 &  [-0.04, -0.11] &  0.13 &  43 \\
 Q2 &  238 &  0.487  &  0.08 &  [-0.04, -0.41] &  0.04 &  134\\
 Q3 &  238 &  0.195  &  0.69 &  [-0.02, -1.65] &  0.00 &  32 \\
 \hline
 C  &  254 &  5.235  &  0.06 &  [0.00, 0.00]   &  0.26 &  45 \\
 Q1 &  254 &  13.286 &  0.06 &  [-0.04, -0.11] &  0.66 &  17 \\
 Q2 &  254 &  1.435  &  0.05 &  [-0.01, -0.40] &  0.23 &  124\\
 Q3 &  254 &  0.360  &  0.61 &  [0.08, -1.91]  &  0.02 &  159\\
 \hline
 C  &  273 &  11.000 &  0.11 &  [0.00, 0.00]   &  0.12 &  26 \\
 Q1 &  273 &  11.549 &  0.06 &  [-0.03, -0.12] &  0.44 &  29 \\
 Q2 &  273 &  1.057  &  0.12 &  [0.09, -0.40]  &  0.20 &  114\\
 Q3 &  273 &  0.286  &  0.50 &  [0.97, -1.58]  &  0.03 &  146\\
 \hline
 C  &  290 &  5.248  &  0.10 &  [0.00, 0.00]   &  0.18 &  28 \\
 Q1 &  290 &  5.669  &  0.06 &  [-0.06, -0.09] &  0.19 &  14 \\
 QN &  290 &  0.165  &  0.10 &  [0.03, -0.14]  &  0.03 &  15 \\
 Q2 &  290 &  0.530  &  0.08 &  [0.04, -0.45]  &  0.05 &  109\\
 Q3 &  290 &  0.251  &  0.70 &  [0.07, -1.53]  &  0.06 &  76 \\
 \hline
 C  &  351 &  1.000  &  0.10 &  [0.00, 0.00]   &  0.12 &  96 \\
 Q1 &  351 &  6.448  &  0.06 &  [-0.07, -0.16] &  0.30 &  17 \\
 QN &  351 &  3.811  &  0.07 &  [0.01, -0.10]  &  0.18 &  13 \\
 Q2 &  351 &  0.248  &  0.24 &  [-0.04, -0.51] &  0.05 &  108\\
 Q3 &  351 &  0.283  &  0.48 &  [0.08, -1.31]  &  0.03 &  66 \\
\end{talltblr}
\end{table*}

A common method of reconstructing VLBI observations is via a geometrical model-fit, as shown, for example, in \cite{Roelofs23}, \cite{Paraschos24a}, \cite{Paraschos24c}, and \cite{Kouch25}.
With this morphological representation, one can associate characteristics of specific model fitted components to jet measurables.
In this work, we represented the jet structure of BL Lac with circular Gaussian components.
With this choice, the number of degrees of freedom was reduced, facilitating the model convergence to a unique solution.
Our choice of using the forward-modelling software \texttt{eht-imaging} \citep{Chael16} was motivated by the need to extract the polarisation signature of individual jet features (which more traditional software like \texttt{difmap} do not provide).
A more detailed description of our approach is discussed in \cite{Paraschos24a} and  \cite{Paraschos24c}.

\section{Numerical setup}\label{app:num}

\begin{table}
    \centering
    \begin{threeparttable}
        \caption{BL Lac jet parameters used throughout the work}
        \label{tab:values}
        \begin{tabular}{c|c}
            \toprule
            Observable & Measurement \\
            \midrule
            \midrule
            $D_\textrm{L}$          & 320.9\,Mpc\tnote{a}                 \\
            z                       & 0.069\tnote{a}                      \\
            $\Gamma_\textrm{beam}$  & 4.5\tnote{b}                        \\
            $M_\textrm{mag}$        & 4.7\tnote{b}                        \\
            $B_\textrm{core}$       & 0.2\,G\tnote{c}                     \\
            r                       & 1.3\tnote{d}                        \\
            $M_\bullet$             & $1.7\times10^{8}\,M_\odot$\tnote{e} \\
            \bottomrule
        \end{tabular}

        \begin{tablenotes}
            \small
            \item[a] \cite{sbarufatti05}
            \item[b] \cite{Cohen15}
            \item[c] \cite{OSullivan09}
            \item[d] \cite{Kramer21}
            \item[e] \citep{Woo02}
        \end{tablenotes}
    \end{threeparttable}
\end{table}

We modelled the polarised synchrotron emission from relativistic jets using three-dimensional special relativistic magnetohydrodynamic (RMHD) simulations computed with the \textsc{PLUTO} code \cite{Mignone07}. 
The system evolves the conservation laws
\begin{equation}
\partial_t \mathcal{U}^k + \partial_i \mathcal{T}^{ik} = 0;
\end{equation}
here $\mathcal{U}$ and $\mathcal{T}$ are the state vector of conservative variables and the flux tensor, respectively.
The simulations were performed in dimensionless units, with physical scaling applied in post-processing to recover characteristic length, time, and energy scales. 
Synthetic full-Stokes ($I$, $Q$, $U$, $V$) emission maps were generated via polarised radiative transfer using \textsc{RADMC-3D} \citep{dull}, including synchrotron emissivity and absorption as well as Faraday rotation and conversion \citep{Nick2018, MacDonald21}.

Our modelling framework follows the methodology introduced in \cite{Kramer21}, where non-thermal electron populations are inferred from thermal plasma quantities through scaling relations, enabling the computation of synchrotron emission in post-processing. 
We followed the same physical scaling as described in~\cite{Kramer24} Sect.~2.3, as well as the initial conditions of the electron energy distribution and magnetic field morphology. 
More precisely, the jet was initialised with a density of $0.1$ code units, corresponding to $1.67\times10^{-23}\,\mathrm{g\,cm^{-3}}$, and injected at a velocity of $0.26\,c$, yielding a kinetic energy density of $\sim2\times10^{-3}\,\mathrm{erg\,cm^{-3}}$. 
The adopted normalised length scale of $0.04\,\mathrm{pc}$ places our simulations in the parsec-scale jet regime.
While hybrid fluid-particle simulations incorporating Lagrangian tracer particles provide a more self-consistent treatment of particle acceleration and radiative losses \citep{Vaidya18, Kramer24}, we adopted here a reduced RMHD-only approach motivated by recent findings that the inclusion of Lagrangian particles does not significantly alter the large-scale morphology or polarisation properties of the resulting synchrotron emission for parsec-scale jets. 
In particular, \cite{Kramer24} demonstrate that while particle-based treatments refine spectral evolution and cooling effects, the global intensity and polarisation structure remain largely consistent with those obtained from standard RMHD post-processing~\citep{Kramer21}. 
This justifies the use of computationally less expensive RMHD simulations for the present parameter study.
Furthermore, although SSC is generally believed to dominate the X-ray and $\gamma$-ray emission of BL Lac objects, particularly within the compact radio and millimetre core on sub-parsec to parsec scales, our simulations probe the parsec-to-kiloparsec jet where synchrotron emission is expected to remain the dominant radiative process. 
To first order, including SSC would primarily modify the high-energy electron distribution through additional radiative cooling, thereby changing the relative brightness and spectral colouring of compact emitting regions. 
In contrast, $\psi$ orientation is expected to remain largely governed by the projected magnetic-field geometry and Faraday rotation. 
Consequently, SSC is anticipated to affect the intensity and colour maps more strongly than the polarisation-vector orientations, except in regions where SSC cooling changes the dominant polarised emitting volume along the line of sight.

The linear polarisation properties are characterised through the Stokes parameters $Q$ and $U$, from which we derive the fractional linear polarisation and $\psi$. 
The polarisation angle is defined as
\begin{equation}
\psi = \frac{1}{2} \arctan\left(\frac{U}{Q}\right),
\end{equation}
and traces the orientation of the projected magnetic field on the plane of the sky. 
For optically thin synchrotron emission, $\psi$ is perpendicular to the projected magnetic field direction.
We estimated the uncertainty of $\psi$ by assuming additive Gaussian, zero-mean, uncorrelated thermal noise with equal root mean square in Stokes $Q$ and $U$, which, through standard error propagation, yields $\sigma_\chi \approx \tfrac{1}{2}\,\sigma_P/P$. 
The uncertainty was computed per pixel using the local polarised intensity $P=\sqrt{Q^2+U^2}$ and summarised by the median and percentile range across the map.

We adapted the numerical setup to reproduce the physical properties of the blazar BL Lac. 
The source was modelled at a luminosity distance of $D_\textrm{L} = 320.9\,\mathrm{Mpc}$ ($z=0.069$;~\citealt{sbarufatti05}), with an bulk Lorentz factor of $\Gamma_\textrm{beam}\sim 4.5$ and a bulk flow characterised by a magnetosonic Mach number $M_\textrm{mag} \approx 4.7$~\citep{Cohen15}. 
All values are summarised in Table~\ref{tab:values}.
Although jets are believed to launch as strongly magnetised outflows~\citep{Tavecchio14, Cohen14}, magnetic energy is expected to be progressively converted into kinetic and thermal energy through magnetic reconnection and jet acceleration~\citep{Walg25}.
We therefore assumed approximate equipartition between magnetic and kinetic (or thermal) energy~\citep{Kramer21}, representing a physically motivated intermediate regime in which the magnetic field still shapes the synchrotron polarisation properties, while the jet dynamics remain predominantly kinetic and free of strong magnetically driven instabilities~\citep{Walg25}.
The specific value for the core magnetic field used in our simulation is of the order of $B_\textrm{core}\sim0.2$\,G \citep{Kramer21}, matching the reported value in \cite{OSullivan09}; the ratio of kinetic to magnetic energy is $r\sim1.3$.

The resulting synthetic emission maps were convolved with a two-dimensional Gaussian function to match the angular resolution of the VLBA at 43\,GHz observations. 
At this resolution, the simulated emission exhibits a predominantly single-sign polarisation structure, consistent with observations of BL Lac. 
Within our modelling framework, this behaviour is best reproduced by a jet configuration dominated by an underlying poloidal magnetic field, in agreement with the polarisation morphology trends identified in \cite{Kramer21}.

\section{Lepton-hadron ratio exploration} \label{app:load}

\begin{figure}
	\centering
	  \includegraphics[trim={4cm 0cm 5.3cm 0cm} , clip, width=0.48\textwidth]{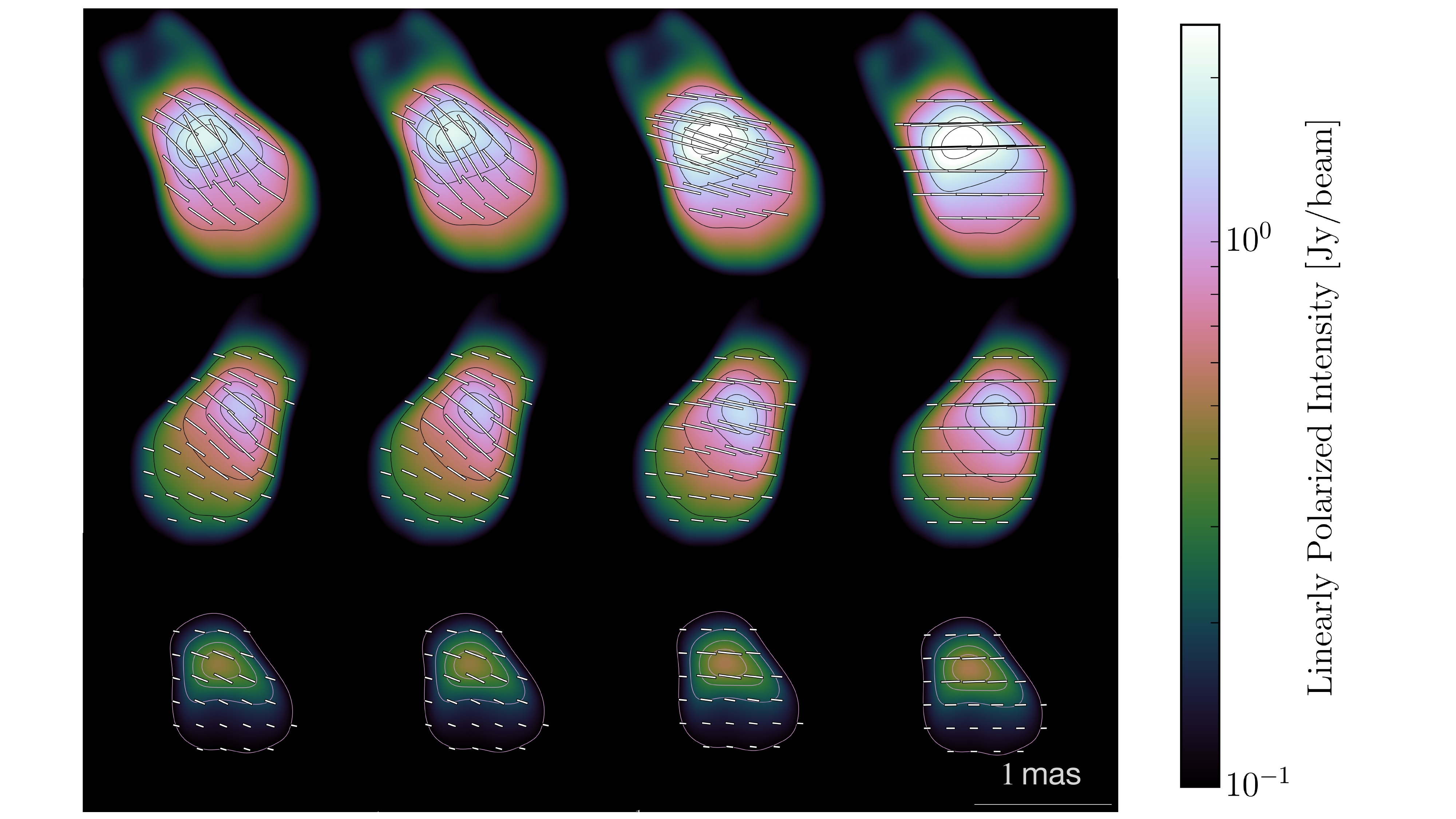}
	\caption{
        2D snapshots of ray-traced 3D RMHD jet simulations. 
        The colour represents linearly polarised intensity in Jansky per beam, overplotted with $\psi$ (white ticks) and total intensity as contours. The columns correspond from left to right to $n=1000, 100, 1$, and $0.01$.
        A clear trend can be identified: when protons are dominating ($n=1000, 100$), $\psi$ are more affected by the orientation of the jet.
        At $n=1$, this dependence is more subtle, whereas in the positron dominated regime $\psi$ remain unaffected by the jet orientation.
            }
	\label{fig:SimP} 
\end{figure}

\begin{figure} 
	\centering
	  \includegraphics[width=0.5\textwidth]{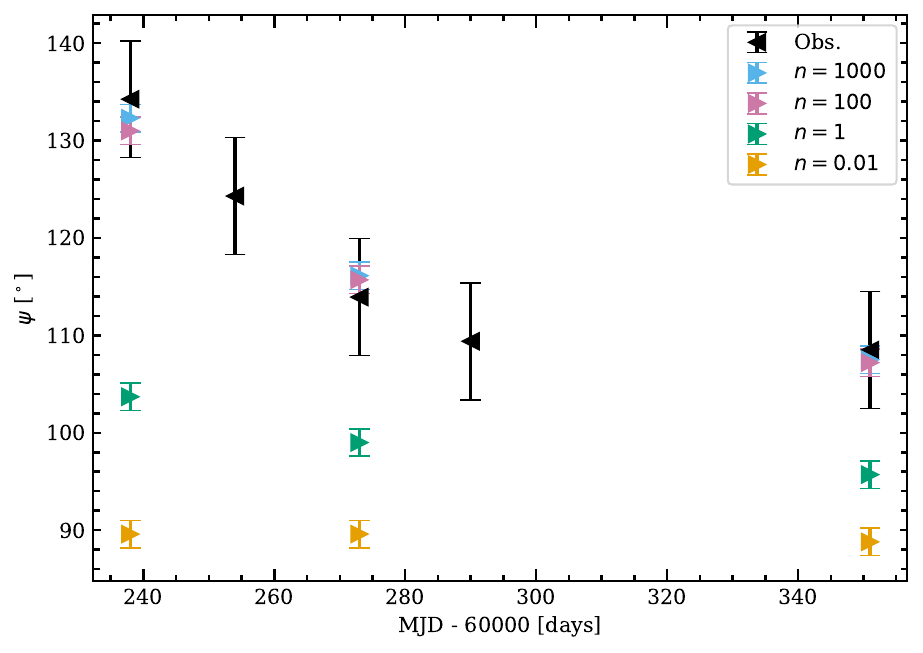}
	\caption{
        Comparison between observed and simulated $\psi$ for different $n$ realisations.
        Similar to Fig.~\ref{fig:SimP}, in the proton dominated regime, the simulated $\psi$ reproduce the observations most faithfully.
            }
	\label{fig:SimE} 
\end{figure}

\begin{table*}[ht]
    \centering
    \caption{Simulated $\psi$ across $n$ and time}
    \label{tab:angle_data}
    \begin{tabular}{l cccc}
        \toprule
        Time & \multicolumn{4}{c}{$n$} \\
        \cmidrule(lr){2-5}
        MJD-60000 & $n = 0.01$ & $n = 1$ & $n = 100$ & $n = 1000$ \\
        \midrule
        \midrule
        238 & 89.6$^\circ\pm$1.4$^\circ$ & 103.7$^\circ\pm$1.4$^\circ$ & 131.0$^\circ\pm$1.4$^\circ$ & 132.3$^\circ\pm1.4^\circ$  \\
        273 & 89.6$^\circ\pm$1.4$^\circ$ & 99.0$^\circ\pm$1.4$^\circ$  & 115.7$^\circ\pm$1.4$^\circ$ & 116.1$^\circ\pm1.4^\circ$  \\
        351 & 88.8$^\circ\pm$1.4$^\circ$ & 95.7$^\circ\pm$1.4$^\circ$  & 107.2$^\circ\pm$1.4$^\circ$ & 107.5$^\circ\pm1.4^\circ$  \\
        \bottomrule
    \end{tabular}
\end{table*}

One of the main parameters that characterise our RMHD simulations is the proton to positron ration $n$.
By comparing our simulations to our observations, we were able to constrain the value of $n$, as shown in Figs.~\ref{fig:SimP} and ~\ref{fig:SimE}.
We find that higher proton values ($n=1000, 100$) reproduce the observed jet flux density and $\psi$ most faithfully, whereas lower proton values ($n=1, 0.01$) produce less pronounced changes both in flux density and $\psi$ rotations.
The corresponding values for $\psi$ and $n$ for three different snapshots are displayed in Table~\ref{tab:angle_data}.
The relatively narrow distribution (see Tab.~\ref{tab:angle_data}) reflects the nearly uniform polarised signal-to-noise ratio across the modelled jet after excluding low-polarisation regions where $\psi$ is poorly constrained.

\section{Proton -- neutrino association}
\label{app:neutrino}

Here we investigate whether the parsec-scale region analysed in this work can act as an efficient neutrino production site.
Throughout we adopted the source parameters of Table~\ref{tab:values} and identified the emission region with component Q2 (Table~\ref{tab:angle_data}). 
For a FWHM of $\sim 0.08$\,mas and an angular scale of $1.36$\,pc\,mas$^{-1}$ (corresponding to $D_\mathrm{L} =
320.9$\,Mpc and $z = 0.069$; \citealt{sbarufatti05}), the comoving radius of the region is $R' \approx 1.7\times10^{17}$\,cm. 
For $\Gamma_\mathrm{beam} = 4.5$ \citep{Cohen15} and a viewing angle $\theta \approx 5\degr$--$8\degr$ (Sect.~\ref{sec:obs}; \citealt{Casadio21}), the Doppler factor is $\delta = [\Gamma(1-\beta\cos\theta)]^{-1} \approx 6$--$8$; we adopted $\delta = 7$.

We note that within our RMHD framework, the protons constitute a cold fluid component: they carry inertia and enter the radiative transfer only through the charge-asymmetry term of the Faraday rotation coefficient \citep{MacDonald21}, and no non-thermal proton population is assumed in producing the simulated $\psi$. 
For the purpose of the present estimate, we therefore adopted the most favourable assumptions, namely that the entire kinetic energy density of the region resides in relativistic protons following a power law with index $\alpha_\mathrm{p} = 2.0$, which is both the best-fit value inferred for TXS\,0506$+$056 \citep{Reimer19} and the most optimistic proton-distribution case of \citet{Bhuyan26}. 
All fluxes derived below are consequently upper limits.

The energy budget is fixed by the equipartition normalisation of our simulations (Table~\ref{tab:values}): with $u'_B = B_\mathrm{core}^2/8\pi\approx 1.6\times10^{-3}$\,erg\,cm$^{-3}$ and $r = 1.3$, the proton energy density is bounded by $u'_\mathrm{p} \approx r\,u'_B \approx 2.1\times10^{-3}$\,erg\,cm$^{-3}$, independently of the value of $n$; the plasma parameter sets the composition of the flow, not its total energy content. 
The proton kinetic luminosity for the jet is given by the following relation \citep{Celotti08}:
\begin{equation}
  L_\mathrm{p} = \pi R'^2 c\, \Gamma^2 u'_\mathrm{p}
  \approx 1\times10^{44}\,\mathrm{erg\,s^{-1}}.
  \label{eq:Lp}
\end{equation}
For a black-hole mass of $M_\bullet \approx 1.7\times10^{8}\, M_\odot$ \citep{Woo02}, the Eddington luminosity is $L_\mathrm{Edd} = 1.26\times10^{38}\,(M_\bullet/M_\odot)\,\mathrm{erg\,s^{-1}} \approx 2.1\times10^{46}$\,erg\,s$^{-1}$, such that $L_\mathrm{p}/L_\mathrm{Edd} \approx 0.5\%$. 
A proton-dominated composition with $n \geq 100$ is therefore energetically unproblematic.

Even under these favourable energetic conditions, the emission region analysed here is not expected to be an efficient neutrino producer. 
BL Lac lacks the luminous external photon fields (broad-line region, accretion disc) that provide dense photomeson targets in more powerful blazars, and its X-ray emission is of synchrotron self-Compton origin \citep{Liodakis25}. 
It, therefore, belongs to the class of low-luminosity, BL Lac-type AGN for which \citet{Murase14} find that external radiation fields are negligible and photomeson production in the jet is inefficient, with the associated neutrino output peaking well above the multi-TeV range of candidate neutrino blazars such as TXS\,0506$+$056. 
This is consistent with \citet{Reimer19}, who show that even TXS\,0506$+$056 itself requires a dense, external soft X-ray target, which is absent in BL Lac, to account for its neutrino emission. 
The baryon loading inferred in Appendix~\ref{app:load} is thus a necessary but not sufficient condition for efficient neutrino production: a detectable signal would additionally require either a much more compact emission region closer to the central engine, or an additional particle-acceleration channel such as magnetic reconnection, as suggested by the multizone RMHD modelling of \citet{Bhuyan26}.

\end{appendix}

\end{document}